\documentclass[12pt,preprint]{aastex}

\def\iso#1#2{\mbox{${}^{#2}{\rm #1}$}}
\def\he#1{\iso{He}{#1}}
\def\li#1{\iso{Li}{#1}}
\def\c1#1{\iso{C}{1#1}}
\def\n1#1{\iso{N}{1#1}}
\def\o1#1{\iso{O}{1#1}}

\def\beq{\begin{equation}}
\def\eeq{\end{equation}}
\def\beqar{\begin{eqnarray}}
\def\eeqar{\end{eqnarray}}

\def\pref#1{(\ref{#1})}

\def\ga{\mathrel{\mathpalette\fun >}}
\def\fun#1#2{\lower3.6pt\vbox{\baselineskip0pt\lineskip.9pt
  \ialign{$\mathsurround=0pt#1\hfil##\hfil$\crcr#2\crcr\sim\crcr}}}

\begin{document}

\title{Probing Primordial and Pre-Galactic Lithium with High Velocity Clouds}

\author{Tijana Prodanovi\'{c} and Brian D. Fields} 
 
\affil{Center for Theoretical Astrophysics,
Department of Astronomy, University of Illinois,
Urbana, IL 61801}

\begin{abstract}

The pre-Galactic abundance of lithium
offers a unique window into non-thermal cosmological
processes.
The primordial Li abundance is guaranteed
to be present and probes big bang nucleosynthesis (BBN),
while an additional Li component
is likely to have been produced by
cosmic rays accelerated in large scale structure formation.
Pre-Galactic Li currently
can only be observed in low metallicity Galactic halo stars,
but abundance measurements are plagued with
systematic uncertainties due to modeling of stellar
atmospheres and convection.
We propose a new site for measuring
pre-Galactic Li:  low-metallicity, high-velocity clouds (HVCs)
which are likely to be extragalactic gas accreted onto 
the Milky Way, and which already have been found to have
deuterium abundances consistent with primordial.
A Li observation in such an HVC would provide the first 
extragalactic Li measurement, and could shed new light on the apparent 
discrepancy between BBN predictions and halo star Li abundance
determinations. Furthermore, HVC Li could
at the same time test for the
presence of non-primordial Li due to cosmic rays.
The observability of elemental and isotopic
Li abundances is discussed, and candidate sites identified.

\end{abstract}

\section{Introduction}

The primordial lithium abundance currently presents a pressing
cosmological conundrum.
The recent {\em WMAP} determination of the cosmic baryon density
\citep{wmap}, combined
with big bang nucleosynthesis theory,
tightly predicts the primordial \li7 abundance \citep{cfo},
but Li measurements in halo stars 
give values lower than this by factors $\ga 2$. 
Several possibilities for this discrepancy
--observational systematics,
stellar destruction,
BBN nuclear uncertainties, new physics--are
discussed in 
\citet{cfo04} 
and references therein.
Until this discrepancy is resolved, 
it challenges the otherwise spectacular agreement 
among {\em WMAP}, BBN, and deuterium \citep[and now \he4;][]{cfos} abundances.

The Li problem becomes even worse when one realizes that there is 
likely to be an {\em additional pre-Galactic source of lithium}
which would have arose during the formation of the Local
Group.  The baryonic matter is almost certain to have
undergone shocks, during mergers and during
accretion onto dark matter potentials;  diffusive
shock acceleration would have led to a population of
relativistic ``structure formation cosmic rays'' 
\citep[SFCRs; see, e.g.,][]{miniati,kwlsh,rkhj}.
These cosmic rays in turn produce both Li isotopes
via $\alpha \alpha \rightarrow \li{6,7}$ reactions
\citep{si,montmerle,prf}.  
This Li component
should add to the halo star Li content, and thus
observed halo star pre-Galactic Li should be corrected downward
for them to get to the primordial 
value.\footnote{Note that this pre-Galactic SFCR Li component is in
addition to any Li produced by normal Galactic cosmic
rays \citep{rbofn},
and differs from that component in that it would be
{\em independent} of halo star metallicity. }
But since observations already give low Li, the problem gets even worse.

To date, halo stars are the only site suitable for observations of
pre-Galactic Li, and have proven a very powerful
tool for studies both of cosmology and of cosmic rays.
But given that the observations are dominated
by systematic errors 
\citep{rbofn,bonifacio}, 
it is critical to identify other independent
sites in which pre-Galactic Li can be measured.
Recently \citet{zl} pointed out that
observations of highly redshifted
($z \sim 500$) lines from  cosmic Li recombination
can be used to probe the Li abundance at these
very early epochs. 
This method could prove very powerful, but
is not yet available.  In the meantime, in this paper we propose
a new site that is currently accessible.

A way to independently 
test the pre-Galactic Li abundance is to look at high velocity clouds 
(hereafter HVC).
These are gas which is falling onto our Galaxy, 
and the lowest metallicity clouds have
a metallicity of about 10\% of solar.
These low-metallicity HVCs thus 
should have a mostly 
pre-Galactic composition, with 
a small contamination from the Galaxy.
Moreover, these cold clouds are free of the possibility of thermonuclear
depletion, which complicates the interpretation of halo star
Li abundances.

Thus measuring Li in HVCs would provide an important test of the Li problem:
if the measurement is consistent with the WMAP+BBN Li abundance
(i.e., at that level or above),  
it would indicate that low Li measured in halo stars is a convection problem, 
or if measurement is below the WMAP result it would indicate that Li problem
is more severe and requires more radical solutions. Also, measurement of Li
in HVCs would test the significance of SFCR contribution to Li production.

\section{High Velocity Clouds}

Clouds of neutral hydrogen HI that significantly depart from the normal 
Galactic rotation, i.e. that have velocities with respect to the
local standard of rest $|v_{\rm LSR}| \ga 90 \ \rm km/s$, are called the
High Velocity Clouds \citep{wvw}. Both positive and negative velocities
are observed, however the sign of their radial velocity 
does not directly imply that their full space
motion is either away or towards the Galactic plane. 
Although
determination of their distances is very uncertain, 
the limits can go up to
tens of kpc \citep{wakker01}. HVCs contain heavy elements
and exhibit a wide range of metallicities,
which in some clouds can be as low as 1/10 of 
solar \citep{wakker01}. 

A few models for their nature have been proposed.
Some HVCs may be of Galactic origin, e.g., in the Galactic fountain model
\citep{sf,bregman}, while others may be extragalactic \citep{oort,blitz,bb},
resulting from the accretion of gas that was left over from the
formation of the Galaxy.
HVCs consistent with Galactic origin would have normal metallicities
as opposed to those with extragalactic origin which are expected to have
lower metallicities \citep{wvw}. 
There is also another type of object, like the Magellanic Stream, which 
represent material that was stripped from satellite galaxies.

When measuring abundances it is crucial
to know the dust content of the HVC,
in order to correct for the depletion
onto dust; 
this effect is known to be very large for local interstellar
abundances \citep{ss}.
The effect of dust is such that it ``hides'' some fraction
of the present abundance and thus introduces non-negligible upward
corrections to the
observations of gas-phase abundances.
In particular, a gas-phase observation of Li is
always a lower bound to the true abundance.
Searches for the dust in HVCs 
\citep{wb,bck,fong}
give negative results, indicating either a dust content
much lower than in low-velocity HI clouds  or
that the dust is very cold \citep{wvw}. Also, \citet{tripp}
found recently 
that HVC Complex C has little or no dust,
based on the iron abundance.
We note that in another class of low-metallicity objects,
the QSO absorption systems, 
there is also evidence that dust depletion is
small, at least in some systems \citep{lrdp}.

Thus, HVCs with low metallicities and little dust, like Complex C 
\citep{tripp}, are very promising
sites for testing the pre-Galactic lithium. Complex C would be particularly
suitable for this measurement since \citet{sembach} have already measured
deuterium abundance there and found it to be consistent with the primordial
abundance inferred form the WMAP. 
Complex C is observed to have low metallicity, 
although   
there are also
indications that its origin might be from the material that was tidally 
stripped from satellite galaxy, like in the case of the 
Magellanic Stream \citep{tripp}. 
In that case, Complex C might not be as pristine
as one would want it to be in order to test the pre-Galactic lithium,
however it is still the most promising candidate for such a task.

\section{Lithium in HVC:  Expectations}

We will take the point of view
that low-metallicity HVCs consist of infalling extragalactic
(i.e., intragroup) matter, with some
admixture of galactic material responsible for the
nonzero metallicity.
We thus expect the HVC lithium
to consist of at least two components:
(1) primordial  \li7, plus (2)
some amount of \li6 and \li7 from galactic processes;
it is also likely that there is a third
component due to SFCRs.
The Galactic Li sources are Galactic cosmic rays 
\citep[which makes \li7 
and are the only
Galactic source of \li6 ; see][]{sw,elisa,fo99},
 and
other sources of \li7: the supernova neutrino process 
\citep[e.g.,][]{whhh}
and low-mass giant stars \citep{sb}. 
In models of the Galactic chemical evolution of Li, 
both the Galactic cosmic-ray
Li components and the supernova component scale linearly
with metallicity \citep{rbofn}.
The evolution of stellar Li is more complex \citep{romano}
but is only important at the highest metallicities
($\ga 10^{-0.8}$ solar), 
and to a rough approximation
also scales linearly with metallicity.
Of course, it is unclear whether the Galactic contribution to
HVC Li should be taken as a diluted form of the
solar component, or as the predicted value
at the HVC metallicity, but as long as the Galactic component
scales linearly with metal content, these two results
should be the same.

Thus the total (pre-Galactic plus Galactic) Li content in
a HVC would depend on the cloud metallicity, and the pre-Galactic
component should be more dominant the lower the metallicity.
One would thus expect to find 
\beqar
 {\rm Li}_{\rm HVC} & \ga & 
 \li7_{\rm p}+\frac{Z}{Z_\odot}\left[ \left( \li7_{\odot}-\li7_{\rm p} \right)
   + \li6_{\odot} \right]  \\
\label{eq:Li_CC}
                & \ga & 7 \times 10^{-10}
\label{eq:limit}
\eeqar
where the notation ${\rm Li} \equiv {\rm Li/H} = n_{\rm Li}/n_{\rm H}$ 
represents the lithium abundance.
 The primordial lithium abundance is given as
 $\li7_{\rm p}$ \citep{cfo}, 
while the solar abundances were taken from
\citet{ag}.
The final, numerical value is that appropriate for
the HVC Complex C \citep{sembach},
which has $Z = Z_\odot/6$ as determined from the
oxygen abundance,
as described in the next section.

\section{Lithium in HVC:  Observability}

Although measuring lithium in the high-velocity clouds would be a way to 
test (and possibly resolve!) the lithium problem and structure formation
cosmic rays, the question is whether this measurement can realistically be 
made. 
Lithium measurements in diffuse gas are particularly difficult because of
the low abundance
and hence small column density. For example, local
ISM observations typically find a
Li column of order $\sim 10^9 \ \rm cm^{-2}$ \citep{kfl}. Thus
to compensate for the low column density and make a successful Li observation
in a cloud of gas one needs to look toward a very bright background object.
In the case of the local ISM,
\citet{kfl} exploited bright stars ($m_V \sim 1-6$) 
to successfully observe diffuse Li, and even to resolve isotopic
lithium abundances using high-dispersion spectra.  
 
Measuring lithium in the HVCs would resemble the ISM measurements in the sense
that both systems contain diffuse, gas-phase Li.
However, the observed HI column in, e.g., the HVC Complex C (towards
the QSO PG 1259+593) is 
$N({\rm HI}) \approx 10^{20} \ \rm cm^{-2}$  \citep{sembach}.
This indicates that the Li column
can be $\ga 10^{10} \ \rm cm^{-2}$;
indeed, eq.~\pref{eq:Li_CC} gives
$N(\rm Li) = 7 \times 10^{10}\ \rm cm^{-2}$
for hydrogen column of $N({\rm HI})= 10^{20} \ \rm cm^{-2}$.
Thus with respect to the
column density HVCs are more favorable sites for measuring Li than the ISM.
On the other hand, local ISM Li measurements can exploit
nearby bright stars,
while for HVC measurements one would have to observe toward an 
extragalactic
object. In that case the brightest candidates are QSOs,
of which the brightest are $m_V \sim 15$, 
about $10^4$ times dimmer than stars used in the ISM measurements.
Finally, one would have to worry about the presence of dust in the
high-velocity clouds, 
but \citet{tripp} found
elemental abundances
which imply that Complex C 
contains little or no dust. 
On the other hand,
depletion onto dust is a significant effect for the ISM Li measurements. 
This is the main reason
why expected  Li column in HVC Complex C ($\sim 10^{10}\ \rm cm^{-2}$)
is so much bigger than the ones reported by  \citet{kfl}.
 
Thus we see that observing Li in a HVC is more challenging
than in the ISM, but the measurement is an important one
and is not beyond the reach of current instruments.
Although it would be very interesting and important to resolve isotopic
lithium abundances in the HVC using a high-dispersion
spectrum, the first step should more realistically be
to obtain an {\em elemental} Li abundance, using a low-dispersion spectrum.
An elemental Li abundance would still provide
important answers about the lithium problem and 
possibly give a valuable insight 
into population of cosmic rays that originate from the large-scale structure
formation.

To get the sense of the observability of elemental
Li, consider the 
\citet{kfl} observations of the ISM lithium,
where {\em isotopic} lithium abundances
were successfully measured and resolved. For example, 
the Li colum density towards the Per X star ($m_{\rm v,*} \approx 6$), is
$N({\rm Li}) \sim 5 \times 10^{9} \ \rm cm^{-2}$, which is about 10 times
lower then the expected column of the elemental Li in the Complex C
towards QSO PG 1259+593 ($m_{\rm v,\rm QSO} \approx 15$). However, the star 
used 
in the ISM Li measurement has about 4000 times larger flux than the quasar
that could be used in the HVC Li measurement. Thus, for the same 
exposure time and spectral 
resolution that was used in \citet{kfl} ISM measurement of 
\li6, the HVC Li observability would be about 300 times lower; that is,
a similar {\em isotopic} measurement
would require that much larger an exposure time. 
The ISM Li isotopes were measured with exposure time of
about $100 \ {\rm ksec}$, so measuring Li isotopes
in HVCs does not seem feasible at present.

However, \citet{kfl} used a 2.7 m telescope for their ISM-Li observation.
Thus if one would to use a 10 m telescope to observe Li in HVC Complex C
this would increase the observability by a factor of about 14, i.e. 
the HVC Li exposure time would now be about 20 times higher compared to the
\citet{kfl} ISM-Li measurement.
This is still quite a chalenge in terms of a reasonable exposure time.
It is important to note that \citet{kfl} measurements were made
with impressive spectral resolution.
However, much lower
resolution would be quite sufficient for measuring 
the {\em elemental} lithium
abundance, as in the first
measurements of elemental Li
in the local ISM \citep{tc,vdb}. 
Thus, by having a spectral resolution that is about a factor
of 6 lower than the one obtained by \citet{kfl}, the exposure time
needed for elemental lithium measurement in the Complex C would be
about $300 \ {\rm ksec}$.
Therefore, even though this is just a crude estimate,
we believe that, although challenging, elemental lithium is reasonably 
observable
in a suitable HVC sightline, such as that towards QSO PG 1259+593.

Thus, we strongly urge that Li be measured in one or more
low-metallicity HVCs, since 
any detection, at or above the level given in equation (\ref{eq:limit}),
 would be of profound importance, especially for the Big Bang Nucleosynthesis
(BBN).

\section{Discussion}

Given the pressing matter of the primordial lithium problem,
namely the discrepancy between the BBN prediction of primordial Li
and the Li observed in low metallicity halo stars, it is crucial
to find new ways of testing this problem. 
In this work we identify an overlooked site where primordial
and pre-Galactic lithium abundances can be tested: low-metallicity
High Velocity Clouds. Although there remain important open questions
about their origin, a number of HVCs were observed to have metalicities
of order 1/10 of solar, indicating that their origin is from the accretion
of gas that was left over from the formation of the Galaxy. 
These clouds, mainly pre-Galactic in composition, are free from
the systematic uncertainties that abundance measurements in halo stars
suffer from due to modeling of stellar atmospheres and convection.
Thus, HVCs are well-suited for testing the lithium problem. 

In this work we address the observability of extragalactic lithium in HVCs,
and conclude that elemental lithium abundance is reasonably observable,
at a level comparable to the
first pionnering measurements of elemental lithium in the local
ISM \citep{tc,vdb}. 
We predict a lower limit to the lithium abundance that is expected
to be observed in low-metallicity HVC Complex C towards the QSO PG 1259+593.

Measuring at or above this limit would be consistent with BBN prediction
of primordial Li abundance and would thus indicate that the solution to
the lithium problem should be found in the stellar modeling. 
Moreover, this measurement would also be a valuable test of additional
sources of pre-Galactic lithium, like structure-formation cosmic rays.
Since the Galactic contribution in eq.~\pref{eq:Li_CC} is
about the same as primordial, a measurement above this
level would
indicate the presence
of additional source of Li (from the presence of dust it always
follows that $\rm Li_{HVC} \ge Li_{obs}$). 
The value in eq.~\pref{eq:Li_CC} includes the
Galactic contribution, which is essentially
``guaranteed.''
In addition,
SFCRs should provide an additional Li source,
particularly if the HVCs really are intragroup gas which has
been exposed to the Local Group SFCR flux.
Recently, \citet{prf} used a model-independent way to constrain the
SFCR-Li abundance range, which by using (\ref{eq:limit})
comes to be about $0.4-5.6$ of the Galactic HVC lithium component.
Thus, if Li in HVCs was found to be sufficiently above
the primordial level, the excess over the Galactic
contribution could be  attributed 
to SFCRs, which would then give us
more insight into this CR population. This way we could limit
the level of contamination of ISM-Li with SFCR-made Li which could
possibly find its way into our Galaxy through the in falling HVCs.
 
However, we stress that measuring  lithium below 
the level in eq.~(\ref{eq:Li_CC}) is also
not excluded, in which case the already existing lithium problem would
become more severe.
Granted, one would
then be able to argue that this just indicates that
there is more dust than it was assumed at first, however one would then have 
to explain
why would lithium be more affected by dust than some other elements which 
indicate
low presence of dust \citep{tripp}.

Another valuable insight we would gain by looking for lithium in
high-velocity clouds  would be
related to the already observed deuterium-- this would be the first primitive
system with both D and Li and thus we could compare their ratio to the BBN and
ISM values.
Finally, if Li isotopic information became available,
one could obtain a more robust separation of
cosmic-ray and BBN components. 

Finally we conclude that measuring even just elemental lithium in 
low-metallicity high-velocity clouds is of great importance, for
it may hold a key to the resolution of the lithium problem, 
and because it will give us a great insight into any additional sources
of pre-Galactic lithium.

\acknowledgments
This material is based upon work supported by the National Science
Foundation under Grant No.~AST-0092939.

\end{document}